\documentclass[12pt,preprint]{aastex}

\received{3484}

\revised{--}

\accepted{--}

\cpright{}{2009}

\slugcomment{To be submitted to ApJ Letters}

\shorttitle{Quadrature Observations of an EUV Wave}

\shortauthors{Patsourakos \& Vourlidas}

\begin{document}

%%%%%

%%%%%

\title{'EUV Waves' are Waves: First Quadrature Observations of an EUV Wave from \textsl{STEREO\/}}

\author{Spiros Patsourakos}

\affil{George Mason University, Fairfax, VA 22030, USA} \email{spiros.patsourakos@nrl.navy.mil}

\author{Angelos Vourlidas}

\affil{Code 7663, Naval Research Laboratory, Washington, DC 20375, USA}

\email{vourlidas@nrl.navy.mil}

\begin{abstract}

  The nature of CME-associated low corona propagating disturbances,
  'EUV waves', has been controversial since their discovery by EIT on
  \textit{SOHO\/}. The low cadence, single viewpoint EUV images and
  the lack of simultaneous inner corona white light observations has
  hindered the resolution of the debate on whether they are true
  waves or just projections of the expanding CME. The operation of the
  twin EUV imagers and inner corona coronagraphs aboard
  \textsl{STEREO\/} has improved the situation dramatically. During
  early 2009, the \textsl{STEREO\/} Ahead (STA) and Behind (STB)
  spacecraft observed the Sun in quadrature having an $\approx
  90^\circ$ angular separation. An EUV wave and CME erupted from
  active region 11012, on February 13, when the region was exactly at
  the limb for STA and hence at disk center for STB. The
  \textit{STEREO\/} observations capture the development of a CME and
  its accompanying EUV wave not only with high cadence but also in
  quadrature. The resulting unprecentented dataset allowed us to
  separate the CME structures from the EUV wave signatures and to
  determine without doubt the true nature of the wave. It is a
  fast-mode MHD wave after all!
\end{abstract}

\keywords{Sun: coronal mass ejections (CMEs)}

\section{Introduction}\label{intro}

An important discovery of \textit{SOHO\/}/EIT (Delaboudini{\`e}re et
al. 1995) was the detection of large-scale EUV disturbances traveling
over significant fractions of the solar disk (e.g., Moses et al. 1997;
Thompson et al. 1998, 1999). These EUV waves emanate from flaring
active regions (ARs) but are strongly associated with coronal mass
ejections (CMEs) onsets (e.g., Biesecker et al. 2002; Patsourakos et
al. 2009). Despite the observations of hundreds of EUV waves over a
full solar cycle, their origin is still strongly debated. A rather
obvious mechanism is a fast-mode MHD wave triggered by the eruption
(e.g., Thompson et al. 1999; Wang 2000; Wu et al. 2001; Ofman and
Thompson 2002; Vr{\v s}nak et al. 2002; Warmuth 2007).  This
interpretation accounts for their association with H$\alpha$ Moreton
waves, their low average speeds (a few hundred $\mathrm{ km
  {s}^{-1}}$;  Long et al.  2008; 
Veronig et al. 2008; Gopalswamy et
al. 2008 for the latest  \textit{STEREO\/}  results), and is the expected plasma behavior after a sudden energy
release (e.g., a flare and/or CME). However, expanding EUV dimmings
are often observed at the wake of EUV waves and sometimes develop
'stationary' fronts which could, in principle, pose problems to a wave
interepretation.  Several authors have thus suggested that EIT waves
are the footprints or the low coronal extensions of the associated
CMEs, and thus are 'pseudo-waves' (e.g., Delann{\'e}e 2000;
Chen et al. 2002; Attrill et al. 2007; Delann{\'e}e et
al. 2008). See also the  review by Warmuth (2007) and 
Patsourakos et al. (2009) for  a 
compilation of \textit{STEREO\/} observational tests for the various wave theories
and their comparison with actual \textit{STEREO\/} observations.

The main reason for the lingering controversy is the lack of
observations with appropriate cadence, and field of view (FOV)
coverage to allow separation between the various facets of the CME and
of the wave.  EUV waves are better observed when their source region
is close to disk center, which allows monitoring of their propagation
over large areas of the solar disk.  On the other hand, CMEs are
better observed off-limb or close to limb, which allows to track their
low-coronal radial and lateral evolution.  Clearly, the single
viewpoint \textit{SOHO\/} observations could address
either the wave or the CME onset but never {\it both} of them at the
{\it same} time. Significant confusion on the nature of the
propagating features associated with the EUV waves has also been
caused by the the relatively low-cadence ($\approx$ 12 minutes) of the
EIT observations. Finally, the lack of an inner White-Light
coronagraph (WLC) on \textit{SOHO\/}, hindered comparisons of simultaneous
EUV images of waves and WLC images of the associated CMEs.

Obviously, the optimal observing configuration for solving the EUV
wave problem are simultaneous EUV-coronagraph observations in {\it
  quadrature}. This was not possible until the launch of the \textit{STEREO\/}
mission in late 2006 (Kaiser et al 2008).  By early 2009, the two
 spacecraft reached a separation of $\approx 90^\circ$, ideal for EUV wave observations.
%The two STEREO
%spacecraft, SC A and SC B, drift apart at a
%rate of $\approx 45^\circ$ per year.
%Despite the very low solar
%activity, an 'EUV wave'-producing active region crossed the solar disk
%in February, 2009.

Here we present the first quadrature observations of an EUV wave. It
emanated from an active region at disk center as viewed from STB but
located at the limb as viewed from STA. Moreover, we have EUV
images at a higher cadence than the \textit{SOHO\/} ones and WLC coverage of the inner corona
(\S~\ref{sec_overview}). With this unique dataset we were able to
simultaneously follow the early evolution of the EUV wave and the CME
at quadrature (\S~\ref{obs}). It was rather straightforward to
determine that the EUV wave is indeed a real MHD wave and not a
pseudo-wave (\S~\ref{kinematics} - \S~\ref{modeling}).

\section{Overview of the CME-Wave Observations}\label{sec_overview}

We use EUV and total brightness WLC images from the Extreme
Ultraviolet Imaging Telescope (EUVI; Wuelser et al. 2004) and the COR1
coronagraph (Thompson et al. 2003) respectively of the Sun-Earth
Connection Coronal and Heliospheric Investigation (SECCHI; Howard et
al. 2008) instrument suite. EUVI observes the entire solar disk and
the corona up to 1.4 $R_{\odot}$. We use images from the 171 and 195
\AA\ (hereafter 171 and 195) channels and our EUVI observations have a
cadence of 2.5 minutes (5 in STB) in 171 and 10 minutes in 195. The
COR1 coronagraph observes the corona in  1.5-4 $R_{\odot}$ with a
10-minute cadence.
%SC B images are taken 19 s after the corresponding
%SC A ones to allow for the different light transit times to the two
%STEREO SC, i.e. they correspond to the {\it same} time on the Sun.

The event took place on February 13, 2009 during a period of
deep solar minimum dominated by quiet Sun. Only a single small
active region 11012 was present over $270^\circ$ of solar longitude. This
unusually 'clean' background helped to unambiguously track
various features associated with the observed wave event at large
distances. A flare-'EUV wave'-CME event originated from this region,
starting at $\approx$ 05:35. The flare was weak (GOES B2.3) and the
corresponding CME was slow ($\approx$ 350 $\mathrm{km\,{s}^{-1} }$ as determined by \textit{CaCTus}, Robbrecht $\&$ 
Berghmans 2004). 
%The active region was located at disk center of SC B and at the east
%limb of SC A.

\begin{figure*}[p]
\centerline{\includegraphics[scale=.85, angle=90]{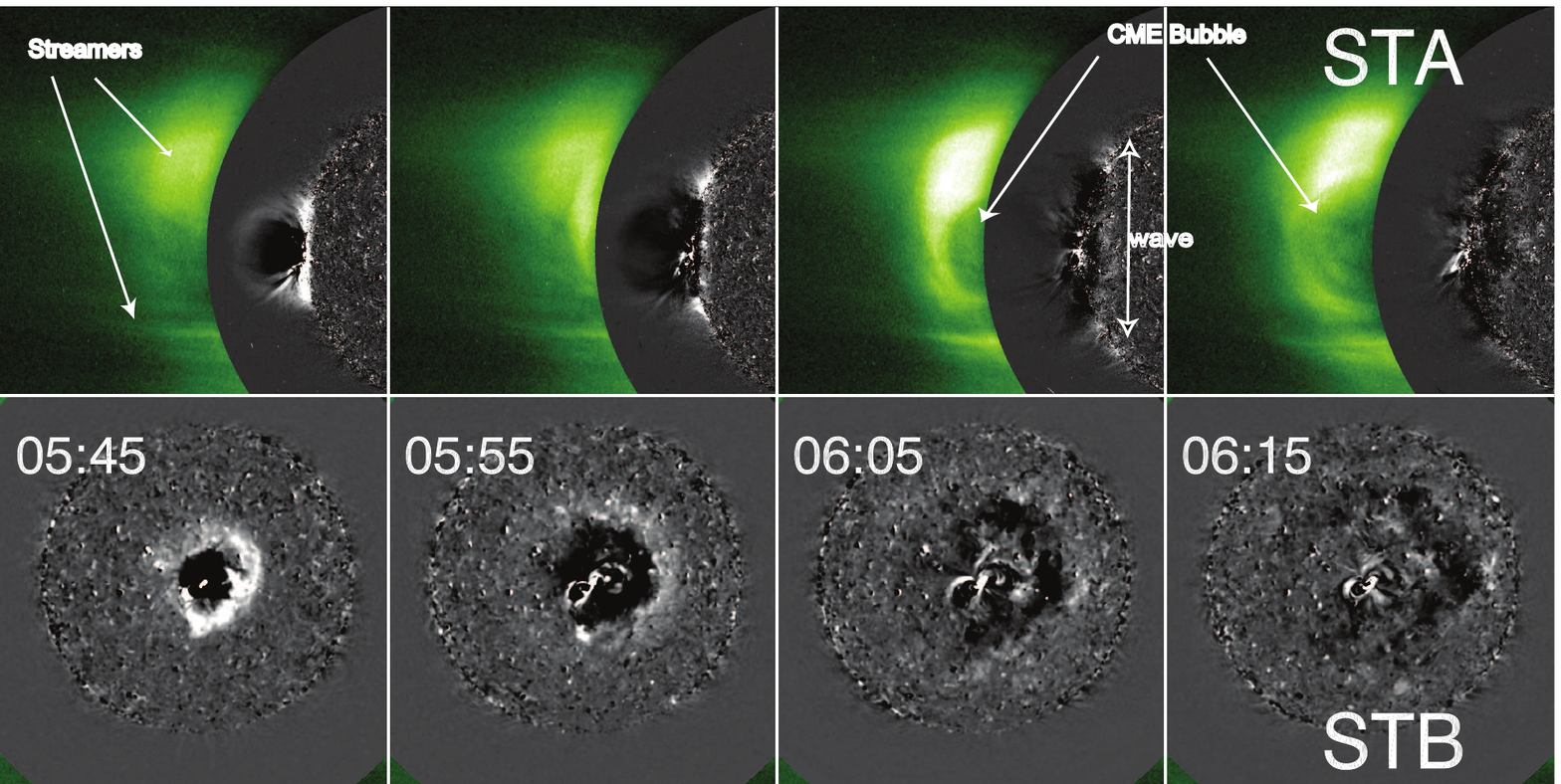}}
\caption{Overview of the quadrature observations of an EUV
  wave. Composite EUVI 195 running difference images (greyscale;
  black(white) implies intensity decrease(increase) respectively) and
  COR1 TB images (intensity increases
  with color from black-white-green) Upper row STA; lower row STB. The
  images for a given instrument were obtained simultaneously on the Sun but the
  time-tags correspond to times for STA.}
\label{overview}
\end{figure*}

Video1.mpg contains STA and STB 195 plain images. Snapshots from the event
in EUVI 195 and COR1 are in Figure~\ref{overview} and the full
development of the wave and the CME can be seen in video2.mpg.  The
195 images are running difference (RD) images (i.e. from each image we
subtract the one 10 minutes earlier).  The COR1 images are total brightness (TB) images. The
FESTIVAL software of Auch{\`e}re et al. (2008)  was used to generate the composite 
EUVI-COR1 images.  The COR1-B images are not shown here
because they do not provide useful information. The CME is a halo in
COR1-B and is only faintly visible late (after 06:55). Starting at
05:35, we observe in EUVI-A a bubble developing both radially and
laterally. The bubble is bounded by streamers in both the north and
south.  When it emerges in the COR1-A FOV at 05:55, it becomes a
rather typical 3-part structure CME. At the same time, the CME pushes
aside streamers on either side. The southern streamer deflection is
especially obvious in Figure~\ref{overview}. Note that the CME cavity
in COR1-A and the EUVI bubble are clearly the same structure (frames
at 05:55 -- 06:25). However, the white light signature of the CME is
much larger than its EUVI counterpart, mostly towards the north.  The
COR1-A CME flanks map very accurately to EUVI dimmings on either side
of the active region and are due to loops deflected by the EUV wave.
By 05:45, the latitudinal extent of the
wave becomes larger than the CME extent. Hence, the STA data alone
reveal very clearly that the CME and the EUV wave are distinct
structures (as also speculated in the review of Harrison 2009 which was
based on pre-\textit{STEREO} data).
with different spatial scales but they also show that
the wave-induced deflections contribute to the width of the white light
CME. It is the later contribution that complicates any CME-wave study
that lacks the data coverage of Figure~\ref{overview}.% without an appropriate definition of the different terms.
%We will return to this important point in 
%  (\S~\ref{conclusions}). 

In EUVI-B, we observe a set of loops that erupt before the wave forms,
which could be a typical pattern in wave formation (e.g. Patsourakos
et al. 2009).  The wave exhibits quasi-circular expansion over most of
the visible disk which is a typical feature of solar minimum EUV waves (e.g.,
Moses et al. 1997; Thompson et al. 1998). There is very little wave
expansion towards the south-east because of the existence of a coronal
hole on the eastern side of the active region. %This is a fortunate
%configuration because in EUVI-A only the western extension of the wave
%is visible. 
The wave becomes more diffuse as it propagates away from
the region and disappears when it reaches the western limb of EUVI-B,
around 06:15. In EUVI-A, the wave extends to about the central
meridian. The latitudinal extention of the wave is the same in both
EUVI-A and B.

\subsection{High-cadence CME-Wave Observations}\label{obs}

\begin{figure*}[p]
\includegraphics[scale=.8]{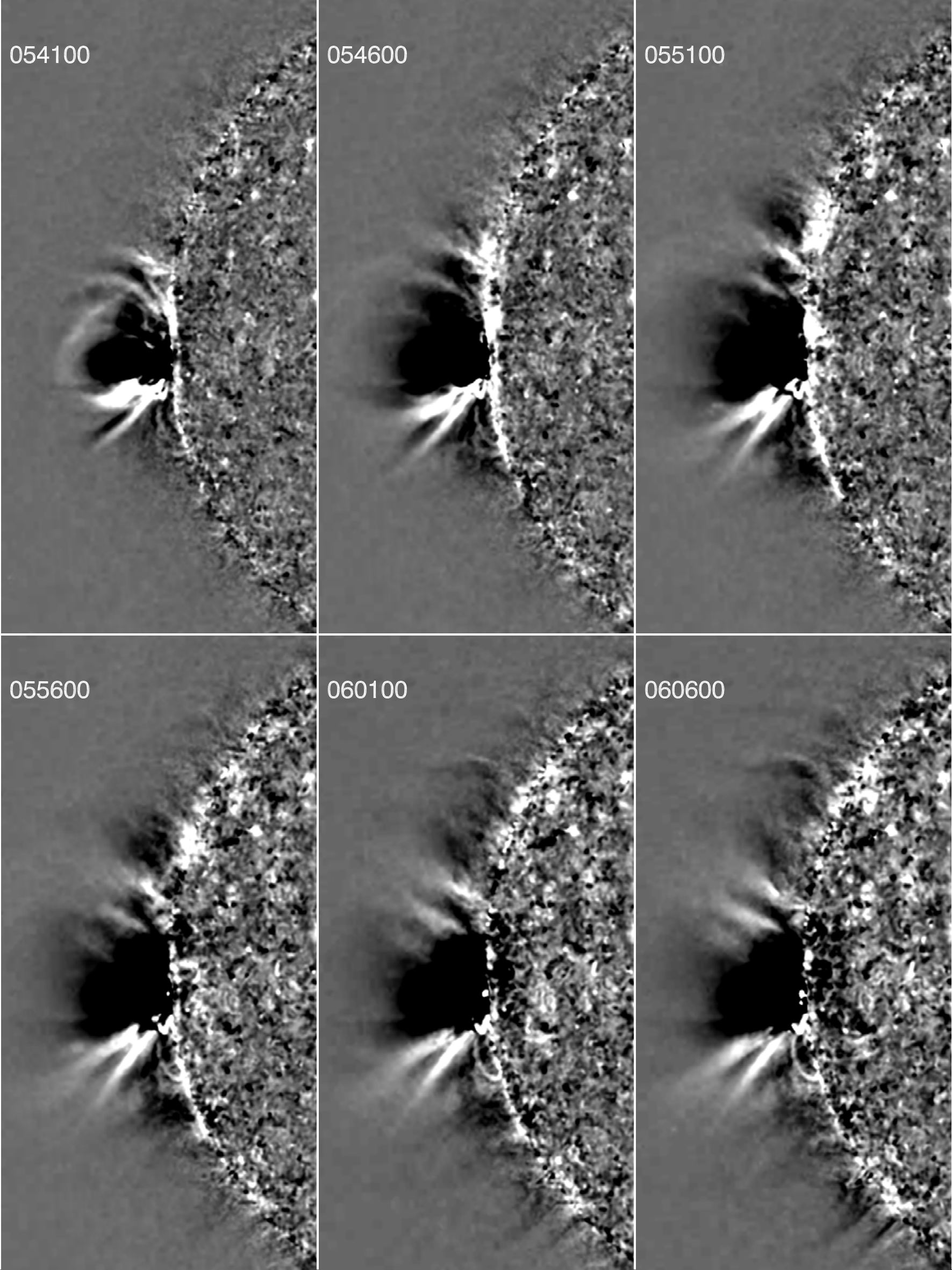}
\caption{Sample snapshots from the 171 SC A image sequence in 10 min
  running difference format.  Black (white) correspond to intensity
  decrease (increase).}
\label{deflection}
\end{figure*}

The high cadence (2.5 min) of the EUVI 171 data allowed us
to understand the nature of the propagating features associated with
the EUV wave. 
%This was not possible before given the low-cadence of
%EIT observations.
We used a 171 EUVI-A movie (video3.mpg) of
wavelet constrast-enhanced images (Stenborg, Vourlidas $\&$
Howard 2008). Sample 10-minute RD frames are given in Figure
\ref{deflection}.

As we saw in the 195 images, a set of low-lying loops, in the shape of
a bubble, starts to slowly rise at $\approx$  05:28. By 05:41, we see the
formation of a dimming at the center of the active region and the
first indications of loop deflections on either side of the expanding
bubble. The deflections appear as black-white pairs in the RD images
and propagate away from the expanding bubble along the north-south
direction. They induce transverse oscillations in coronal structures
at the solar limb. The oscillations dump within 10 minutes and their
maximum amplitude decreases with distance from the source. The outermost
deflected structures seem to match with the latitudinal extent of the
wave as seen on the disk strongly suggesting their association.  These
deflections are likely the off-limb counterpart of the transverse
(kink-like) oscillations seen in active region loops in the wake of an
eruption (e.g., Aschwanden et al. 1999; Nakariakov et al. 1999;
Verwichte, Nakariakov $\&$ Cooper 2005).  The observed off-limb EUV
deflections are not uncommon: they have been observed in EUVI
high cadence movies of eruptions at the limb.  A detailed satistical study of the characteristics
of these deflections (e.g., amplitudes, periods etc) is underway.
Similar deflection
phenomena have been observed with coronagraph in association with CMEs
(e.g., Gosling et al. 1974; Sheeley, Hakala and Wang 2000; Vourlidas
et al. 2003; Liu et al. 2009). The present event showed also evidence
of a streamer deflection in the coronagraph data (e.g., Figure
\ref{overview}). The deflections can only be explained by the passage
of a wave and are therefore a very strong indication that the EUV wave
is indeed a wave.  MHD simulations show ample evidence for deflected
coronal structures once a velocity pulse (i.e. an eruption) is set up
(e.g., Vourlidas et al. 2003; Ofman 2009 for a review).

\subsection{CME-Wave Kinematics}\label{kinematics}

To clarify further the difference in the nature of the CME and the EUV
wave, we performed {\it simultaneous} measurements of the CME and wave
widths. In determining the wave width we followed the method of
Podlachikova $\&$ Berghmans (2005).  We used 195 and 171 STB BD
images where we subtracted a pre-event reference image taken at 05:00.
All images were differentially rotated to the time of the reference
image. The BD images were first projected onto a spherical polar
coordinate system ($\phi$-$r$) with its center on the eruption
site. Data were then discetized on a grid with a $d\phi$=45 $^\circ$
and $dr= 0.075 R_{\odot}$ (or 50 Mm).  We averaged those maps over the
sectors in the NW direction, where the wave was best visible (Figure
\ref{overview}) and obtained radial intensity-ratio profiles. The wave
front location corresponds to the local maxima of these curves. The
wave width error bar was set to $dr$.

For the determination of the CME width we used the wavelet
constrast-enhanced 171 EUVI-A images of the CME bubble.  The latter
was defined as the outermost set of loops which were 'opened' by the
eruption and remained opened. They correspond to the 'deep' dimming
seen in the active region core after the eruption (i.e., the dark
central area of Figure \ref{deflection}).  No deflections were seen
within this area which is consistent with an ejection and not the
passage of a wave. The EUV bubble can be further traced to the COR1
FOV (e.g., the three leftmost panels of Figure \ref{overview}) and was
visible in 171 between $\approx$ 05:36-05:56.  We manually selected a
series of points outlining the bubble and fit them with a circle.  The
radius of the best-fit circle supplied the width of the CME bubble. We
estimated error-bars using the standard deviation of the residuals
between the best-fit circle radius and the distances of the manually
selected points using the best-fit circle center as a
reference. %A sample of the circular fits is given in Figure
% \ref{bubble_171}.  
The above process was also applied to COR1 A images of the CME bubble from 06:10 to 06:35.

\begin{figure*}[p]
\includegraphics[scale=.9]{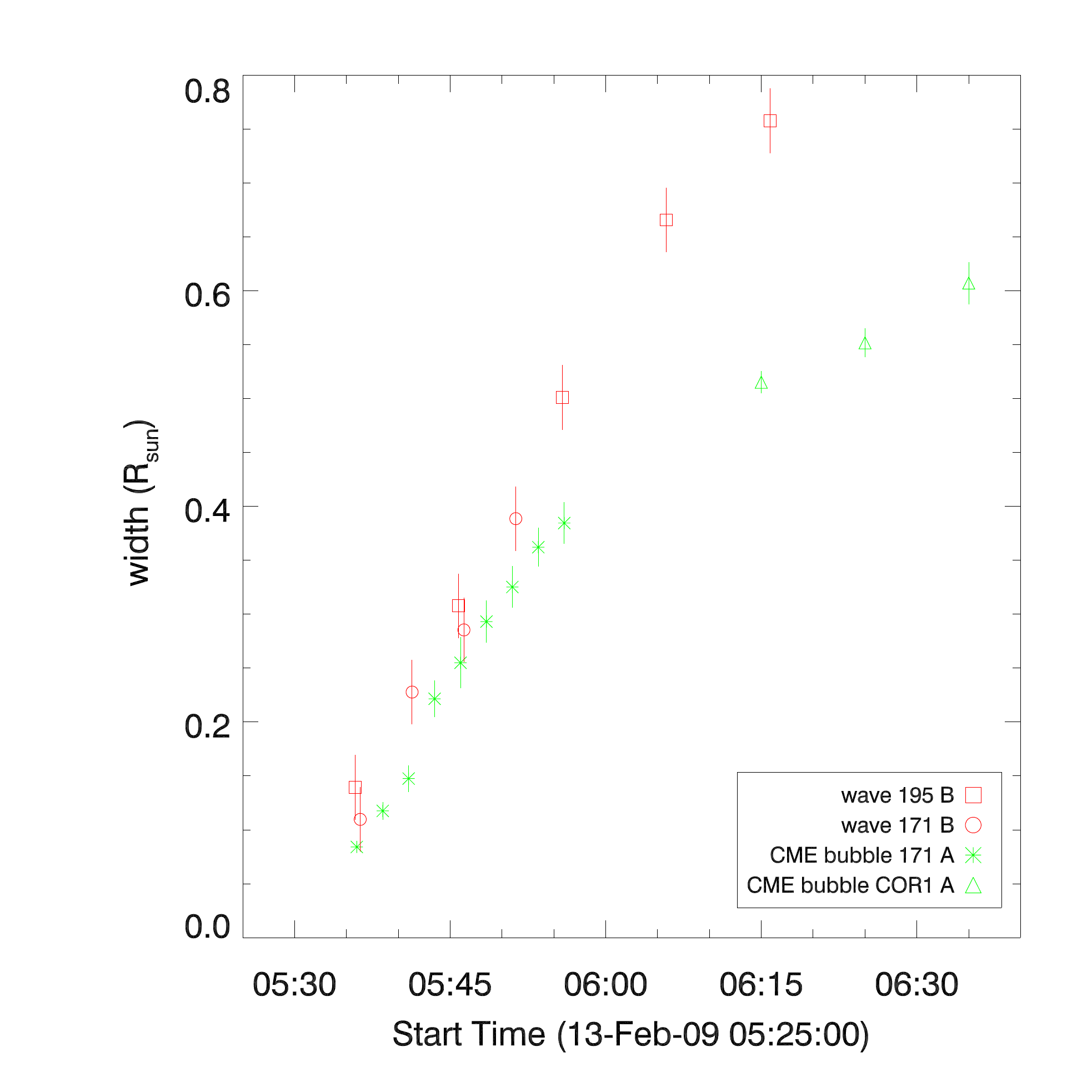}
\caption{Time-evolution of the CME-wave widths from STA and STB respectively. 
See \S~\ref{kinematics}.}
\label{htplot}
\end{figure*}

The CME-wave width measurements are in Figure \ref{htplot}. First note
that the 171 and 195 wave measurements are consistent with each
other. Quadratic fits to the wave width give a linear
expansion speed of $\approx$ 250 $\mathrm{{km}\,{s}^{-1}}$ and an a
decceleration of $\approx$ -25 $\mathrm{{m}\,{s}^{-2}}$, typical
values for EUV waves.  The evolution of the CME width exhibited two
phases: first, a period of strong lateral expansion in the EUVI FOV,
followed by a slower expansion in the COR1 FOV.  The important result
in Figure \ref{htplot} is that while the CME-wave widths track each
other quite closely in the beginning of the event, the wave becomes
significantly wider that the CME after $\approx$ 05:45.
This is in
disagreement with the predictions of pseudo-wave theories which
require that the projected CME width or its low coronal extention to
match the wave width at all times. 

\subsection{3D CME-Wave Modeling}\label{modeling}

Finally, we performed forward modeling of the observed CME and wave
using the simplest instance of the 3D forward model of Thernisien et al. (2006, 
2009)
model; a spherical bubble attached to a conical leg.  The free
parameters of the model were varied until we found a satisfactory
projection of the model into the STA sky plane (see Patsourakos et
al. 2009 for details on the application of this model to EUVI and COR1
data).

\begin{figure*}[p]
\includegraphics[scale=.98]{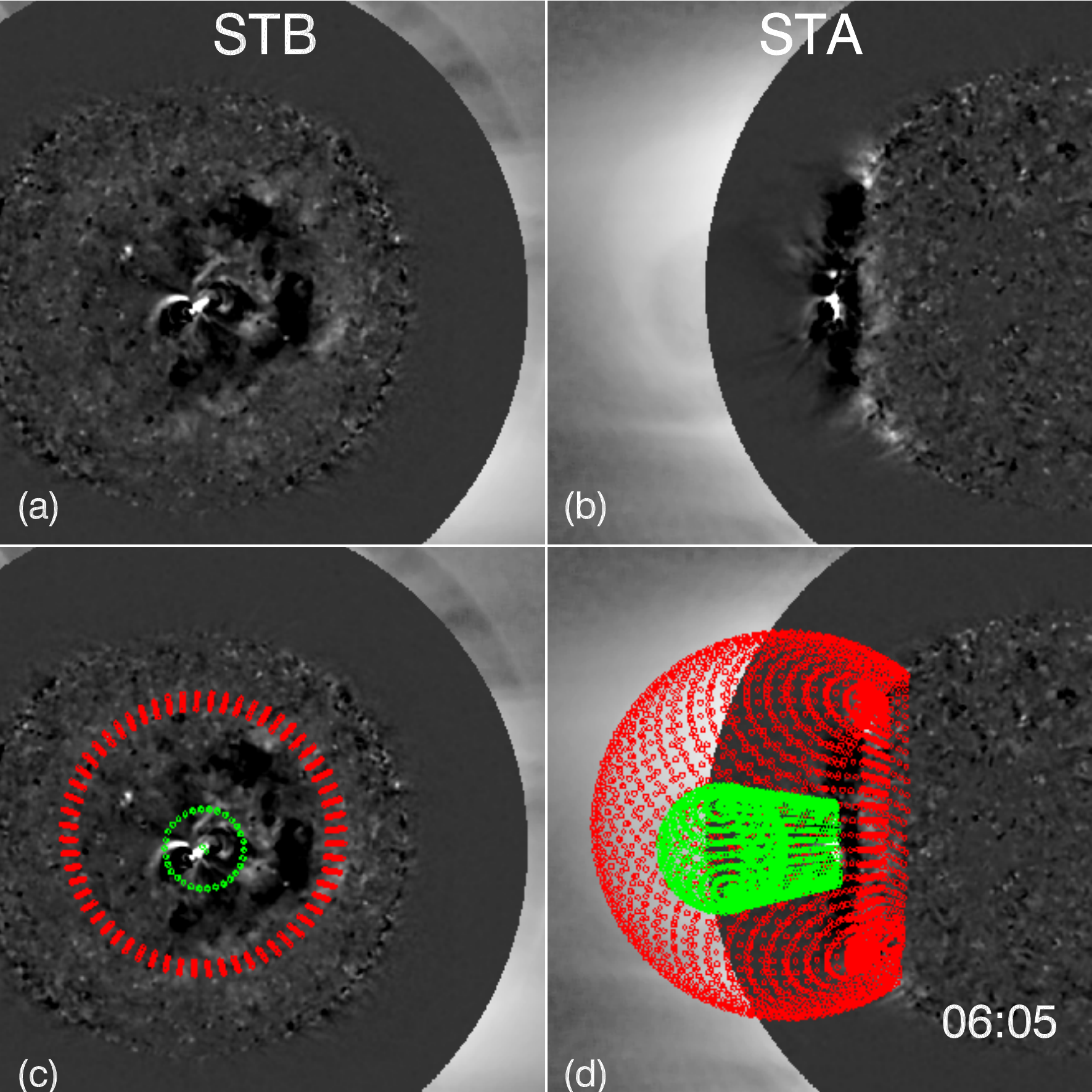}
\caption{Forward modeling of the CME and wave 3D shape in both STEREO
  spacecraft for observations at 06:05. Panels (a) and (b) contain
  composite EUVI 195 RD and COR1 TB images from STB and STA. Panel (d)
  contains the best-fit CME (green wireframe) and wave (red wireframe)
  model determined for STA. Panel (c) has the disk projections of
  these models in STB.}
\label{3dplot}
\end{figure*}

Figure \ref{3dplot} shows the CME-wave modeling for the observations
at 06:05. We selected this time because the wave has covered a
significant part of the visible disk in STB (panel a) while part of
the CME bubble has entered into the COR1-A FOV (panel b).  The model
of the CME bubble (panel d; green wireframe) fits the white light/EUV
cavity rather well with the exception of the rapidly converging legs
of the cavity. This is a limitation of our geometric model. A larger
model was then used to fit the outer boundary of the coronal volume
affected by the eruption (panel d; red wireframe).  The model
encompasses the latitudinal extent of the EUV wave in STA (compare
with panel b) which is comparable to the latitudinal extension of the
off-limb deflected structures (\S~2.3). We note though that the model somehow
overestimates the southward extension of the off-limb volume affected by the
wave. 
Panel (c) contains the disk
projections (in STB) of the models. The wave projection fits
rather well the extent of the wave while the CME projection is smaller
and confined around the deep active region core dimming.  The forward modeling
suggests that the wave and the CME are not concentric with each other or
with the active region center. The wave offset is likely due to the
influence of the coronal hole at the east of the erupting active
region. The CME offset is caused by the westward location of the
erupted loops in the active region core. 

Therefore, the 3D modeling reveals the different scales and nature of
the CME and the wave and provides a straightforward explanation for
the white light extension of the event which is commonly refered to as
the 'white light CME'. It is the latter, rather careless, use of
terminology that seems to be the cause of confusion in EUV wave
studies.

\section{Conclusions}\label{conclusions}

As discussed in \S~\ref{intro}, the exact nature of the EUV waves
(MHD waves or pseudowaves) and their association with CME structures
has been the matter of intense debate since their discovery.  The main
reason was the lack of high cadence comprehensive coverage of the
early development of the wave and of the associated CME. 

A second reason is the careless use of the term 'CME' in a generic way to
variously describe the ejected fluxrope, the full extent of the white
light brightness enhancements and the extent of EUV dimmings in the
low corona. While all of these phenomena are observationally related
to the eruption process, they do not necessarily share the same
physical mechanism, as we have shown here. 

\begin{figure*}[p]
\includegraphics[scale=.90]{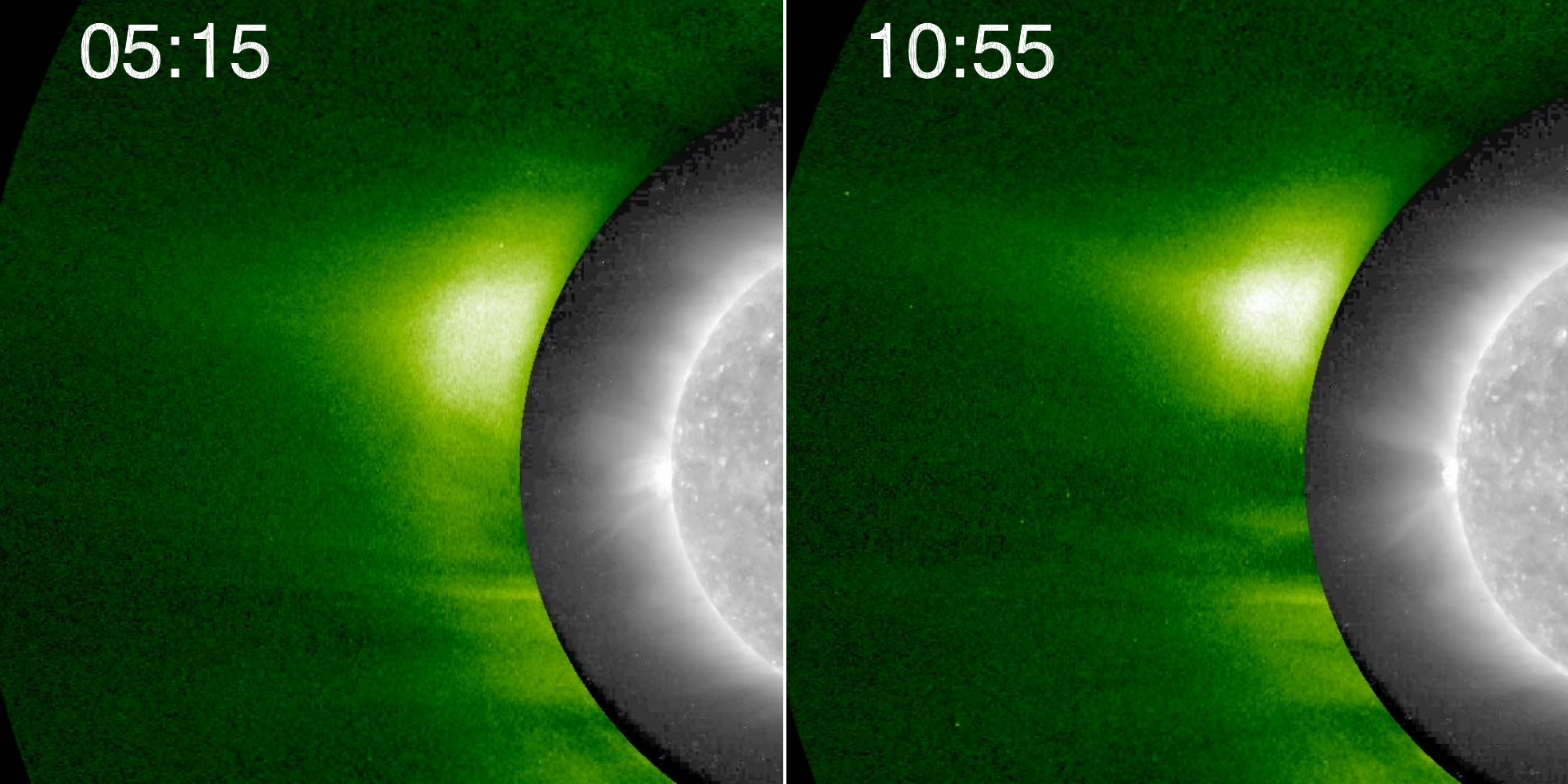}
\caption{Composite of plain 195 and  TB COR1 STA images before and after
  the event (left and right panels, respectively).}
%Examples of circular fittings of the CME bubble in EUVI-A. Purple crosses:
%manually selected points outlining the bubble, white dashes and green box:
%best-fit circle and its center respectively.}
\label{bef_aft}
\end{figure*}

We use the unique SECCHI observations presented above to clarify both
the nature of EUV waves and the CME terminology. For the latter, we
reserve the term 'CME' only for the actual ejected coronal magnetic
structure; the fluxrope from the active region core. This is further
illustrated in Figure \ref{bef_aft}, where it is clear that only the
active region core loops and overlying corona have been removed by the
CME, whereas a much larger coronal volume, including the adjacent
streamers, that participated in the event (or was affected by it)
stayed put after the eruption.  This work shows clearly that our
proper definition leads to a better understanding of the CME and its
effects on the surrounding corona.

The high-cadence, quadrature SECCHI observations of a typical EUV
wave/CME event led us to significantly new insights on the nature and
development of EUV waves and CMEs.  Our main findings are:
\begin{itemize}
\item The CME is born from the transformation of a set of rising loops
  in the core of the active region to a rapidly expanding EUV
  bubble/cavity.
\item The impulsive acceleration of the CME bubble induces deflections
  on progressively remote coronal structures which match with the
  latitudinal extent of the wave.
\item The expanding CME evacuates a significant part of the active
  region corona leading to stanionary dimmings  on the scale of the
  active region.
\item The expanding EUV wave is tracked by a diffuse weak intensity
  enhancement with a traling dimming disturbance in the low corona and
  by deflections of distant (from the active region) streamers higher
  in the corona. 
\item After a few minutes ($\approx$ 15 min), the wave width becomes
  significantly larger than the CME one. The CME width is determined
  by the expanding cavity.
\item 3D modeling of the CME and wave structures showes unambiguously
  that they correspond to different structures: the wave occupies and
  affects a much bigger volume than the CME.
\end{itemize}

{\it All} of the above findings are consistent only with an expanding
fast-mode wave from the site of an impulsive energy release. They are
also consistent with a driven wave (by an expanding CME) and not with
a blast wave (induced by a flare).  These findings, especially the
distant streamer and EUV off-limb structures  deflections are incosistent with the 
notion that 
EUV
waves are pseudo-waves, i.e. the disk projection or the lower coronal
extend of the CME. We must conclude, therefore, that the observed wave
is a true MHD wave. The wave is driven by the expanding CME. The
propagating deflections seen in the off-limb coronal structures and in
the white light streamers higher up serve as a 'smoking-gun' of the
passage of a wave in the corona.

We emphasize here that our discussion applies only to propagating
distrurbances reaching global scales. Such events are typical of solar
minimum conditions, when few active regions are present, and most of
the solar disk is occupied by quiet Sun.  The expanding CME cavity and
stationary dimmings in the active region could very well be accounted
for by the pseudo-wave theories (e.g. Zhukov \& Auch{\`e}re 2004). It could
be well that the observed wave, in the period before wave and CME
start to decouple (i.e. 05:45), was indeed a pseudo-wave.

There are also occassions, particularly under solar maximum conditions
when multiple active regions are present, where EUV dimmings develop
away from the active region and after the wave has passed from these
areas. Those dimmings could originate from reconnections with the
expanding CME fluxrope. So there is no reason to discard those
theories at the moment. Only their application need to be carefully
considered. We hope that the images movies and discusssion in
this work make clear which structure could be the result of a wave and
which cannot.

The SECCHI data used here were produced by an international consortium
of the Naval Research Laboratory (USA), Lockheed Martin Solar and
Astrophysics Lab (USA), NASA Goddard Space Flight Center (USA),
Rutherford Appleton Laboratory (UK), University of Birmingham (UK),
Max$-$Planck$-$Institut for Solar System Research (Germany), Centre
Spatiale de Li\`ege (Belgium), Institut d’ Optique Th\'eorique et
Applique\'e (France), and Institut d’Astrophysique Spatiale (France).

%Add a wave/cME cartoon

%%%%%%%%%%%%%%%%%%%%%%%%%%% Figures %%%%%%%%%%%%%%%%%%%%%%%%%%%%%

%

\end{document}